\documentclass[aps,prl,superscriptaddress,twocolumn]{revtex4-1}
\usepackage{pb-diagram}
\usepackage{pstricks,graphicx,xcolor}

\usepackage{mathrsfs}
\usepackage{latexsym}
\usepackage{amsmath,amsthm,amsopn,amsfonts,amssymb,epsfig,stmaryrd}
\usepackage[active]{srcltx}

\makeatletter

\def\la{{\lambda}}

\def\cal L{{\mathcal L}}

\let\n\noindent

\let\la\lambda
\let\La\Lambda
\let\Om\Omega

\let\ta\theta

\let\rw\rightarrow

\let\y\infty
\let\si\sigma

\def\beq{\begin{equation}}
\def\eeq{\end{equation}}
\newcommand{\ba}{\begin{array}}     \newcommand{\ea}{\end{array}}

\begin{document}

\title{The supersymmetric Ruijsenaars-Schneider model}

\author{O. Blondeau-Fournier}
\email{olivier.b-fournier.1@ulaval.ca}
\affiliation{D\'epartement de physique, de g\'enie physique et
d'optique, Universit\'e Laval,  Qu\'ebec, Canada,  G1V 0A6.}

\author{P. Desrosiers}
\email{patrick.desrosiers.1@ulaval.ca}
\affiliation{D\'epartement de physique, de g\'enie physique et d'optique, Universit\'e Laval,  Qu\'ebec, Canada,  G1V 0A6.}
\affiliation{ CRIUSMQ, 2601 de la Canardi\`ere, Qu\'ebec, Canada, G1J 2G3}



\author{P. Mathieu}
\email{pmathieu@phy.ulaval.ca}
\affiliation{D\'epartement de physique, de g\'enie physique et
d'optique, Universit\'e Laval,  Qu\'ebec, Canada,  G1V 0A6.}


\begin{abstract}
 An integrable supersymmetric generalization of the trigonometric Ruijsenaars-Schneider  model is presented whose symmetry algebra includes  the super Poincar\'e algebra.     
Moreover, its Hamiltonian is showed to be diagonalized by the recently introduced Macdonald superpolynomials.
{Somewhat surprizingly, the consistency of the scalar product forces the discreteness of the Hilbert space.}
\end{abstract}



\maketitle

\n
   {{\bf Introduction.}} In this Letter we resolve a long-standing problem about the existence of a supersymmetric and integrable generalization of the quantum version of the trigonometric Ruijsenaars-Schneider  (tRS) model \cite{RS,RS2}.  

The tRS model is the relativistic generalization of the Calogero-Sutherland (CS) model,    in which $N$ particles interact pairwisely on a circle through a
long-range potential  \cite{CS}. 
``Relativistic'' here means that the model contains a parameter playing the role of the speed of light $c$ and when $c\rightarrow \infty$ it reduces to the CS model (albeit with a rescaling of the coupling constant).   
 This interpretation is further supported by the presence of the Poincar\'e algebra  in the algebraic structure of the model.

The eigenfunctions for the CS Hamiltonian  were found to  be of the form (ground state)$\times$(Jack polynomials) (see \cite{KK} for review). The Jack polynomials (Jacks), fundamental objects  of algebraic combinatorics \cite{Mac} and representation theory \cite{DG}, have found various physical applications recently: quantum fractional Hall states \cite{BH}, singular vectors in CFT \cite{CFT} and AGT-type conjectures \cite{Alba}.

The tRS model is also notorious  {mainly for the importance of its eigenfunctions,  which take the form} (ground state)$\times$(Macdonald polynomials) \cite{KvD}.  
  The Macdonald polynomials (Macs) \cite{Mac} are 
natural generalizations of many basic symmetric functions, including the Jacks. Like the latter, they are relevant in
representation theory \cite{GH} and in various physical contexts: $q$-deformation of the Virasoro algebra \cite{qVir}, the five-dimensional equivalent of the AGT conjecture \cite{5DAGT} and particular solvable probabilistic problems, the Macdonald processes \cite{BC}.

The supersymmetric   generalization of the CS model was discovered  at the beginning of the 1990s  \cite{SUSY}.  The complete understanding of its symmetry algebra and  eigenfunctions  appeared a decade later \cite{DLM12}.   The eigenfunctions are described by the superspace analogues of the Jacks, the Jack superpolynomials  (sJacks), studied in depth in \cite{DLM34}.  Recently, the sJacks were used in    {superconformal field theory} \cite{DLM6} and were shown to possess clustering properties similar to those observed for the quantum fractional Hall states \cite{DLM5}.

Finding the  {supersymmetric} extension of the tRS model has remained an open  problem for two decades.     The rational behind  its difficulty is that the standard techniques of  supersymmetric quantum mechanics \cite{Witten}  
no longer apply in  the relativistic setting.

Here we prove the existence of a generalization of the tRS model that is both supersymmetric and integrable.  The supersymmetry is explicit since the Hamiltonian and the total momentum are generated by fermionic charges, all together forming  the $\mathcal{N}=1$ super Poincar\'e algebra.  We moreover show that the eigenfunctions of the   {supersymmetric tRS (stRS)} model are built  in terms of the Macdonald superpolynomials ({sMacs}).  The   latter  were recently introduced as a superspace generalization of the {Macs}   \cite{BDLM1,BDLM2}.  {Our results thus link the sMacs to a supersymmetric many-body problem.}

\noindent {\bf{The tRS model.}} Before constructing the supersymmetric model, let us recall   important features of the tRS model \cite{RS2,Konno}.  The model involves $N$ bosonic particles   interacting on a ring of length $L$ via a (real positive) coupling constant  ${g}$.   Their dynamics is described by the rapidity variables $\eta_j$ and their  canonical conjugate  $\zeta_j \, (j=1, \ldots, N)$.
The former are chosen in their differential representation,  $\eta_j=-i {\hbar}\partial/\partial \zeta_j$, which guarantees that $[\zeta_j,\eta_k ] = i {\hbar}\delta_{jk}$ and the latter are real  variables.    
We consider all the masses to be identical and equal to $1$ and we set $\hbar=1$.

The interaction between the particles is induced via the ``potential'' functions $V_j = \prod_{k\neq j} h^-(\zeta_{jk})^{1/2}$ and  $W_j = \prod_{k\neq j} h^+(\zeta_{jk})^{1/2}$, where  $ \zeta_{jk}=\zeta_j-\zeta_k$ and   \vspace{-5pt}\beq \vspace{-5pt}
h^\pm({\zeta})=\frac{\sin {\tfrac{\gamma}{2}}({\zeta}\pm g/{c})}
{\sin {\tfrac{\gamma}{2}}(  {\zeta} )}, \qquad {\gamma=2\pi/L.} \eeq
The Hamiltonian  and total momentum   are respectively 
\beq
\begin{split}
\label{HamiltonianRS}  
& H_ {\text{tRS}}={\tfrac{1}{2}\sum_{j }( {V_j} \tau_j^{-1}  W_j +W_j \tau_j V_j) }\\
&
  P_ {\text{tRS}}={\tfrac{i}{2}\sum_{j }( {V_j} \tau_j^{-1} W_j -W_j  \tau_j V_j) }
	\end{split}
\eeq

\vspace{-0.3cm}
\n where $\tau_j=\exp{(i \eta_j/c)}$ is a translation generator, e.g., $(\tau_1 f)(\zeta_1, \ldots, \zeta_N) = f(\zeta_1+1/{c}, \zeta_2, \ldots, \zeta_N)$.  
The Hamiltonian in \eqref{HamiltonianRS} differs from the one presented in \cite{RS2} by the substitution $c \rightarrow ic$. It is the quantum version of the $\text{III}_b$ model in \cite{RIIIb} (further commented in the Conclusion).  The explicit dependance on $c$ can be absorbed in  $\gamma$ after the rescaling $(\zeta_i,\eta_i)\to (\zeta_i/c,c\,\eta_i)$. We can thus set it equal to 1 without loss of generality.

\n Together with the Lorentz boost, $B=- \sum_{i=1}^N \zeta_i$, the Hamiltonian and the momentum form the Poincar\'{e} algebra in 1+1 dimensions {(with $H_\text{tRS}=H, P_\text{tRS}=P$)}: 
\beq\label{PA} 
[H,P]=0,\quad [H,B]=iP, \quad [P,B]=-i H .
\eeq

From now on, it is convenient to change the variables and redefine the parameters as follows \cite{Konno}: 
\beq\label{newvars}
x_j = e^{ i  \gamma \zeta_j}, \quad q=e^{i \gamma}, \quad t=q^g
\eeq
and adopt the Macdonald's notation \cite{Mac}: 
\beq 
A_i(t)=\prod_{j\neq i}\frac{tx_i-x_j}{x_i-x_j} 
\eeq
so that $\prod_{j\neq i} h^{\pm}(\zeta_{ij})=t^{\mp(N-1)/2} A_i(t^{\pm1})$.  The operator $\tau_j$ now represents the $q$-shift operator on functions of the $x_i$, e.g., $(\tau_1 f)(x_1, \ldots, x_N) = f(qx_1, x_2, \ldots, x_N)$.

\n  {{\bf Supersymmetric generalization.}}
We now supersymmetrize the tRS model, {proceeding in five steps.}

\noindent
{\textit{First step}}.  
 We first add fermonic variables to the tRS model.
To each variable $\zeta_j$ (or equivalently, the $x_j$), which is a bosonic degree of freedom,  we associate a fermionic partner  $\theta_j$ ($\ta_j\ta_k=-\ta_k\ta_j$, so that ${\theta_j}^2=0$).   Any  function $\Psi$  depending  upon the bosonic variables $x_j$  and the fermionic variables  $\theta_j$  (to be referred to as a superfunction), and upon the parameters $q$ and $t$,  decomposes as follows:
\beq \label{decompo}
\Psi(x,\theta;q,t) = \sum_{I \subseteq (1,\ldots,N) }\Psi_I(x;q,t) \, \theta_{I} ,\eeq
where  the sum extends  over all  sequences   {of $m$ indices} 
{$ I=(i_1,\ldots,i_m)\;\text{such that}\;  1\leq i_1<\ldots<i_m\leq N$,}
 with $0\leq m\leq N$.  Moreover,  each $\Psi_I$ is a complex valued function  and $\theta_I=\theta_{i_1} \cdots \theta_{i_m}$.  The value of $m$ is called the fermionic degree.   


\n\textit{Second step}. We next specify the  nature of the Hilbert space and in particular, its inner product.
The states are superfunctions $\Psi=\Psi(x,\theta;q,t)$   {that  are periodic under $\zeta_i \rightarrow \zeta_i+L$  and} satisfy $\sum_{I}\int_{{T}}  w_I \, \Psi_I\,\overline{\Psi_I}\,dX <\infty,$
where $ T=\{(x_1,\ldots,x_N)\,|\,x_i\in\mathbb{C},\,|x_i|=1,\forall \,i\}$ and $dX=dx_1/x_1$ $\cdots$ $ dx_N/x_N$.  The ``bar'' operation stands  for the usual complex conjugation which here means $\overline{(x,q,t)}=(x^{-1},q^{-1},t^{-1})$.     
In addition, a non trivial weight functions is introduced: $w_I= t^{m(N-1)/2}\prod_{j\in I} A_j(1/t)$, where $m$ represents the cardinality of $I$ (cf.~below Eq.\ \eqref{decompo}).   
The set of all states forms a  vector space $\mathscr{H}$ that is naturally equipped {with the following} scalar product:
\beq
 \label{scalprod} \langle \Psi|\Phi\rangle:= {\frac{1}{(2\pi i)^N N!}}\sum_{I}\int_{ {T}}  w_I \, \Psi_I(x;q,t)\,\overline{\Phi_I(x;q,t)}\,dX.
\eeq 
Note that  two states of different fermionic degree are orthogonal.
This scalar product reduces to that of the   {sCS}  (resp.\ tRS) model in the non-relativistic   (resp.\   {non}-supersymmetric) limit.    
 {Within the state space  $\mathscr{H}$,
we  focus on the subspace} $\mathscr{H}^{S_N}$ formed by all    {{symmetric states}},    {namely, states that}  are invariant under any simultaneous exchange of pairs of partners $(\zeta_j,\theta_j) \leftrightarrow (\zeta_k,\theta_k)$.

The presence of the weight function $w_I$ makes the calculation of the adjoint operators 
{somewhat subtle}.  For instance,  
  the adjoint of $\theta_i$ is
$\theta_i^\dagger=t^{(N-1)/2}A_i(1/t)\partial_{\theta_i}$. 
Thus, the fermionic variables and their adjoints generate a  {novel and interesting deformation of the} Clifford algebra, $
\{\theta_i,\theta_j^\dagger\}=t^{(N-1)/2}A_i(1/t)\, \delta_{ij}$,
which reduces to  the usual one as $c\to \infty$ (i.e., $t\to 1$).
As a second example, consider the adjoint of the operator $ \tau_i^{-1}$, which  requires the introduction of a projection operator,
\beq 
\pi_I=\prod_{i\in I}\theta_i\partial_{\theta_i}\prod_{j\notin I}(1-\theta_j\partial_{\theta_j})
\quad \Rightarrow\quad 
 \pi_I  (\theta_J)=\theta_{I} \delta_{I J}\,.\eeq
One finds that $
   (\tau_{i}^{-1})^\dagger = \sum_{I\subseteq (1,\ldots,N)}w_I^{-1} \,    { \tau_{i}}  \,w_I\,\pi_I$,
 where 
\vskip-0.7cm
 \begin{equation}{\textstyle
 w_I^{-1} \, \tau_i \, w_I=Z_{(I,i)}\left[\prod_{j\neq i}\frac{(x_i-x_j)(q x_i-tx_j)}{(x_i-tx_j)(q x_i-x_j)}\right]^{\chi(i\in I)}  \!\!\!\!\!\!\!\! \tau_i }
 \end{equation}
with $\chi(\cdot)=1$ if its argument is true and 0 otherwise,  and  
\beq{\textstyle Z_{(I,i)}=\prod_{j\in I, j\neq i} \frac{(q tx_i-x_j)(x_i-x_j)}{(q x_i-x_j)(tx_i-x_j)}.}\eeq
In absence of fermions, $w_I=1$ and $
   (\tau_{i}^{-1})^\dagger=\tau_i$.

{A key point that has not been addressed so far is the positivity of $w_I$ when $q$ and $t$ lies on the unit circle. This discussion is postponed until we complete the formulation of the model.}

\n\textit{Third step}. We introduce the {supersymmetry charge} of the stRS model:
\begin{equation}
Q_{-}= {i} \sum_i\ta_i (a_i-1)\, ,\quad a_i =V_{i}^{-1}\tau_{i}^{-1}W_{i}\,.\label{ai}
\end{equation}
Its adjoint with respect to the scalar product is  
\beq
Q_{-}^\dagger={-i}\sum_i (a^\dagger_i-1)\theta_i^\dagger\, ,\quad a_i^\dagger =W_{i}(\tau_{i}^{-1})^\dagger V_{i}^{-1}\,.\label{ai2}
\eeq
These charges are fermionic.    Moreover, one easily checks that $[a_i,a_j]=0$.  This readily implies that $Q_-^2=0$, which in turn implies that $(Q_-^\dagger)^2=0$. 

The states annihilated by $Q_-$ are called supersymmetric.  For instance, any state of the form $\theta_1\cdots\theta_N \Phi_{(1,\ldots,N)}(x;q,t)$ is supersymmetric.  Most importantly, the ground state $\psi_0$ of the non-supersymmetric tRS model   is also supersymmetric.  This can be understood as follows.  One can show that the ground state wave function is given by $\psi_0=(C_0\Delta_N)^{1/2}$, where  
\beq   
\label{DeltaN}{
\Delta_N= \prod_{ i\neq j}^N \frac{(x_i/x_j;q)_\y}{(tx_i/x_j;q)_\y} 
}
\eeq  
with $(a;q)_n=\prod_{i=0}^{n-1}(1-aq^i)$.  Since  $a_i\psi_0=\psi_0$ for all $i$, one gets that $Q_-(\psi_0)=0$, {while} $Q_-^\dagger \psi_0=0$ immediately follows from the definition. 

Equation \eqref{DeltaN} is problematic in view of the definition of our parameters $q$ and $t$ (cf. eq. \eqref{newvars}). Indeed, $|q|=1$ and this makes the  infinite products of the form $(p;q)_\y$ with $|p|=1$ diverge.
{This issue is ignored for the moment and is reconsidered below in the light of the consequences of imposing the positivity of $w_I$.  }

The two charges $Q_{-}$ and $ Q_{-}^\dagger$ allow us to define the Hamiltonian of the  stRS model:
\beq \label{superHQm}
H=\tfrac{1}{2} \{ Q_-,Q_-^\dagger\} +\varepsilon_+ 
\eeq
where $\varepsilon_+= { - \sum_{i=1}^N\cos[ {\gamma g }(2i-N-1)/2]}$ is a constant introduced for later convenience.  By construction, the Hamiltonian is self-adjoint, its spectrum is bounded from below by ${{\varepsilon_+}}$, and it is supersymmetric: $
[H,Q_-]=[H,Q_-^\dagger]=0\, .$
Moreover, $H$ generalizes both the Hamiltonian of the tRS and the  sCS models.  Indeed, one can check that $H\Psi =H_{\text{{tRS}}}\Psi$ whenever $\Psi$ does not depend on fermionic variables $\theta_i$. Moerover, up to a normalization factor, $\lim_{t\rw 1}H$ is given by ($\beta=g^{-1}$) 
 \beq \nonumber
-\frac{1}{2}\sum_{i=1}^{N} \left(\frac{\partial}{\partial \zeta_i}\right)^2+  \sum_{1\leq
i<j\leq N}\frac{\beta[\beta-1+(\theta_i-\theta_j)(\partial_{\theta_i}-\partial_{\theta_i})]}{ {(2/\gamma)}^2\sin^2({\gamma \zeta_{ij}/2)}},
\eeq
{which is precisely the sCS model's Hamiltonian {\cite{SUSY,DLM12}}.}

\n\textit{Fourth step}.  We introduce another charge:
$
Q_{+}=\sum_i\ta_i (a_i+1)$.
This allows us to define the {momentum} operator:
\beq\label{superPQm}
 {P=\tfrac{1}{2} \{ Q_-, Q_+^\dagger\}+ \varepsilon_-,}
\eeq
where ${\varepsilon_-= \sum_{i=1}^N\sin[ \gamma g (2i-N-1)/2]}$.  In addition 
to $ [H,P]=0$, we have:
\vspace{-0.1cm}
\beq\label{superHPQp}
\begin{gathered}
{\tfrac{-1}{2}}\{Q_+,Q_+^\dagger\} = H+\varepsilon_+, \quad \tfrac{1}{2}\{Q_+,Q_-^\dagger\} = P+\varepsilon_-,\\
\{ Q_\pm,Q_\mp \}=\{ Q_\pm^\dagger,Q_\mp^\dagger \}=0 .\\
\left[Q_{\pm},H\right]=[Q_{\pm}^\dagger,H]=[ Q_\pm,P]=[ Q_\pm^\dagger,P]=0 .
\end{gathered}\eeq 

\vspace{-0.2cm}

\n\textit{Fifth step}.   Here we build the full supersymmetry algebra. By setting
 \vspace{-.1cm}
\beq \begin{gathered}\mathcal{Q}_1=\tfrac{1}{2}(Q_-+Q_-^\dagger),\quad   \mathcal{Q}_2=\tfrac{-1}{2}(Q_++Q_+^\dagger),\\ 
 {\textstyle \mathcal{B}=B+\tfrac{1}{2}\sum_j\theta_j\partial_{\theta_j},}\end{gathered}\eeq 
 (note that these operators are self-adjoint),  we find 
\beq \label{superPoincare}  \begin{gathered} \{\mathcal{Q}_a,\mathcal{Q}_b\}={\sigma_{ab}}H+{(\delta_{ab}-1) \,P}-{\delta_{ab}}\,\varepsilon_+,\\
[\mathcal{Q}_a,\mathcal{B}]=\tfrac{{1}}{2i}{\epsilon_{ab}}\mathcal{Q}_b,\quad [\mathcal{Q}_a,H]=[\mathcal{Q}_a,P]=0,\\
  [H,P]=0,\quad [H,\mathcal{B}]=i P,\quad [P,\mathcal{B}]=-{i} H,\end{gathered} \eeq where  $a,b\in\{1,2\}$, {$\epsilon_{ab}$ is the Levi-Civita symbol, } and $\sigma=\text{diag}(1,-1)$. Eq. \eqref{superPoincare} is  the $\mathcal{N}=1$  {super Poincar\'e} algebra in 1+1 dimensions (see for instance \cite{SA} and references therein).

\n{{\bf Macdonald superpolynomials}.}  
We now solve explicitly the supersymmetric model just constructed.  We concentrate on the space $\mathscr{H}^{S_N}$ of symmetric states and decompose each symmetric $H$- and $P$-eigenstate as $\psi_0 \times f$,   for some symmetric {superfunction} $f=f(x,\theta;q,t)$.

Let us characterize $f$.  For this, we first write the Hamiltonian in \eqref{superHQm} and the momentum in \eqref{superPQm} as
\beq
H=\tfrac{1}{2}(H_{-1}+H_1), \quad P=\tfrac{1}{2}(H_{-1}-H_1),
\eeq 
and factor out the ground-state contribution from $H_{\pm1}$,
\beq 
H_{\pm1} = t^{ \mp N(N-1)/2}\Delta_N^{1/2} {D_{\pm 1}} \Delta_N^{-1/2}.\eeq
 {Thus, $\psi_0\times f$ is an eigenfunction commun to $H$ and $P$ if and only if $f$ is an eigenfunction commun to $D_{-1}$ and $D_{+1}$.  The  latter are obviously supersymmetric since 
$D_{\pm 1}=\{ \hat{Q}_{2\pm1}, \hat{Q}_{3\pm1} \}$, where} 
\beq
\begin{split}
&{\textstyle \hat{Q}_1=\sum_i \theta_i \tau_i^{-1},  \qquad \hat{Q}_2=\sum_i A_i(1/t) \partial_{\theta_i},} \\
& {\textstyle \hat{Q}_3=\sum_{I,i} A_i(t) \, Z_{(I,i)} \, \tau_i  \, \pi_I  \partial_{\theta_i}, \qquad \hat{Q}_4=\sum_i\theta_i.}
\end{split}
\eeq
{One can show that the symmetric eigenfunctions of $D_{{\pm 1}}$  are of the form $\sum_I\theta_I f_I $, where each $f_I$ is a Laurent polynomial  in the variables $x_i$  with coefficients that are rational in $q,t$.  However, whenever $g$ is an eigenfunction of $D_{{\pm 1}}$,  then so is $(x_1\cdots x_N)^{k} g$  for any integer $k$. Thus, the only relevant eigenfunctions are those that are  polynomial (not Laurent-type) in $x_i$ and $\theta_i$. 
Such eigenfunctions for $D_{\pm1}$ were} recently introduced \cite{BDLM1,BDLM2};  they are the Macdonald superpolynomials, sMacs for short.

{Like any symmetric superpolynomial, {the} sMacs are labelled by {superpartitions}.}
 A superpartition, denoted $\La$,  is a pair of partitions $\La=(\la;\mu)$ such that $\la$ is a strictly decreasing partition and $\mu$ is a (regular) non-increasing partition. A superpartition is  said to be of degree $(n|m)$ if $n=|\la|+|\mu|$ and $\la$ has exactly $m$ parts (counting one possible part equal to $0$). 

The sMacs, denoted  $P_\La=P_\La(x,\theta;q,t)$, for superpartitions of degree $(n|m)$ form a basis for the {space} of all symmetric superpolynomials of homogeneous degree $n$ {in the} variables $x$ and fermionic degree $m$ {and with rational coefficients in $q,t$ (considered as independent parameters).}
They are orthogonal w.r.t. 
\beq \langle f|g\rangle':= {\sum_I\int_T \psi_0^2 \,w_I \, f_I(x;q,t) \overline{g_I(x;q,t)} \,dX }.
\eeq
{Note that this scalar product can be rewritten as 
$ \langle f|g\rangle' = \int_T \psi_0^2\,  \overline{ f(x,\theta;q,t)} g(x,\theta;q,t)  dX$, where it is understood that  $\bar \theta_i =\theta_i^\dagger$  and that  $\theta_i^\dagger$ is acting on $ g_I$.
}

{As mentioned above, the   $P_\La$'s are the  symmetric eigenfunctions of $D_{\pm1}$.  More generally, 
$D_{\pm1} (x_1\cdots x_N)^{k}P_\La $ is equal to  $\epsilon_{\La^*+\kappa}(q^{\pm1},t^{\pm1})  (x_1\cdots x_N)^{k}P_\La$, 
where 
$\La^*$ is the partition obtained from the superpartition $\La=(\la;\mu)$ by removing the semi-colon and reordering the parts, $\epsilon_{\La^*}(q,t) = \sum_i q^{\La^*_i}t^{N-i}$,  $k\in\mathbb{Z}$,  and $\kappa$ is the $N$-vector with all components equal to $k$.  Thus, the set of all the states $\Psi_{\La,k}$, such that $\Psi_{\La,k}= (x_1\cdots x_N)^{k}{\psi_0}P_\La$, diagonalizes the  {Hamiltonian} $H$ and momentum $P$.    An orthogonal basis for   $\mathscr{H}^{S_N}$ is then easily formed by making use of
 $\langle \Psi_{\La,k} | \Psi_{\Om,\ell} \rangle   = \langle P_\La | P_{\Omega^\times} \rangle'\propto \delta_{\La  \Om^\times}$ for $k\leq \ell$, where $ \Omega^\times$ is the superpartition obtained  by replacing each element $\Omega_i$ of  $\Omega$ by $\Omega_i-k+\ell$.}


\n{\bf Integrability of the stRS model.}
The proof of integrability relies on the construction of the sMacs in terms of the non-symmetric Macs  \cite{BDLM2}, themselves eigenfunctions of the Cherednik operators. These operators are constructed out of
 the Hecke algebra generators 
\beq
\label{Ti}
T_i=t+\frac{tx_i-x_{i+1}}{x_i-x_{i+1}}(s_i-1), \quad i=1, \ldots, N-1,
\eeq
where the $s_i$ are the elementary transpositions such that $x_i \leftrightarrow x_{i+1}$.  The inverse of $T_i$ reads $T_i^{-1}=t^{-1}-1+t^{-1}T_i$. The Cherednik operators are \cite{Cher}
\beq
Y_i=t^{-N+i}T_i\cdots T_{N-1}\omega T_1^{-1}\cdots T_{i-1}^{-1},
\eeq
for  $i=1, \ldots, N $ and $\omega=s_{N-1} \cdots s_1 \tau_1$. {Importantly, these operators} satisfy $[Y_i,Y_j]=0$ for all $i,j$.
Now, let
\beq
\begin{split}&\Gamma(u;v)=\prod_{i=1}^m (1+ u v Y_i) \prod_{i=m+1}^N (1+u Y_i),\\
&G(u;v)= \sum_{\sigma \in S_N} \sigma\left( \frac{\alpha_1}{\alpha_t} \Gamma(u;v) \frac{\alpha_t}{\alpha_1} \pi_{(1, \dots, m)}  \right),
\end{split}\eeq
where $\alpha_t=\prod_{i<j}^m(tx_i-x_j)$ and the permutation $\sigma \in S_N$ is such that {$(x_i,\theta_i) \mapsto (x_{\sigma(i)}, \theta_{\sigma(i)})$}.  In \cite{BDLM2}, it {was} showed that the generating functions 
{$G(u;1)=\sum_{n=1}^Nu^nD_n$ and $G(u;q)=\sum_{n=1}^Nu^nI_n$}
contain $2N$ independent commuting quantities whose common eigenfunctions are the sMacs of fermionic degree $m$.
The commutativity $[D_i,D_j]=[I_i,I_j]=[D_i,I_i]=0$ follows from {that} of the Cherednik operators. Let us define the operators $D_n$ and $I_n$ with $n\leq -1$ by replacing $Y_i$ by their inverse in $G(u;v)$ (the resulting $2N$ new operators of course are not independent conserved quantities). Since the stRS  {Hamiltonian} is a combination of the $D_i$'s, it follows that $[H,D_i]=[H,I_i]=0$ 
and the integrability of the stRS models is proved, at least in the subspace $\mathscr{H}^{S_N}$.  The extension of this conclusion to the full space $\mathscr{H}$  relies on the generalization of the argument of \cite[App. C]{KvD}.

\n
{{\bf Summing up: main results.}} We have thus succeeded in formulating a supersymmetric version of the tRS model that displays both integrability and super-Poincar\'e invariance.  In addition,   the model  eigenfunctions
have been shown to be the  ground state times the recently-found Macdonald superpolynomials.

\n{
{\bf Back to the scalar product.}} In that regard, that the parameters $q$ and $t$  lie on  the unit circle becomes quite natural. Indeed, the integral version of the {sMacs} scalar product involves a term evaluated at $q, t$ and the other at $1/q,1/t$ \cite{BDLM2}. Rephrased as a quantum mechanical scalar product ($\int \psi^*\psi$), the origin of this inversion must be rooted in complex conjugation, which forces $|q|=|t|=1$.
{This motivated our initial choice for taking the parameter $c$ to be purely imaginary. However, some results appear to be well-defined only when $q,t$ are real and $<1$. This is true, in particular, for the convergence of the expression for the ground-state wave function $\sqrt{\Delta_N}$ and the positivity requirement on the scalar product which forces $w_I\geq 0$.}

{
Now, when $|q|=|t|=1$, the positivity of $w_I$ is no longer automatic. In terms of the original  variables $\zeta_j$,
\beq
w_I=\prod_{j\in I} \prod_{k\neq j}\frac{\sin {\tfrac{\gamma}{2}}({\zeta_{jk}}- g)}
{\sin {\tfrac{\gamma}{2}}(  {\zeta_{jk}} )}.\eeq
If we order the particles such that $\zeta_j>\zeta_k$ if $j>k$, it is not difficult to see that $w_I\geq 0$ requires
\beq\label{basic}
g\leq \zeta_{jk}\leq \frac{2\pi}{\gamma}- g.\eeq 
Here $j,k\in I$ but since we eventually sum over all sectors $I$, the above condition must hold for all $j,k$ (more precisely, with the ordering specified below in eq. \eqref{lattice}). 
Quite amazingly, this is the precise condition required for the classical model to be well-defined \cite{vDV}. 
We stress that it arises here from the consistency of the scalar product in the fermionic sector, a purely supersymmetric feature. 
}

{
In the quantum case, the different states are related by the action of $\tau_i$ which shifts the value of $\zeta_i$ by 1. {The particle configurations are then} in correspondence with the points of a finite lattice. Like for the non-supersymmetric case,
 where the lattice points are delimited by an integral affine alcove \cite{vDV},  the lattice points for which the sRS model is defined are described by restricted superpartitions $\La$ corresponding to the specialization:
\beq\label{lattice}
\zeta_{\si^{-1}(j)}=-\La^*_{j}+(j-1)g,\eeq 
where $\sigma$ is the {smallest}
permutation defined by $\si(\La^a,\La^s)=\La^*$, {the partition obtained by ordering all parts of $\La$.  In particular, the constraints on $\zeta_{N1}$ imply the following restriction on the superpartitions}
\beq
{\La^*_1-\La^*_N\leq M},
\eeq
 where $M$ is an integer defined as
\beq
M=\frac{2\pi}{\gamma}- Ng\qquad \Rightarrow\qquad q^Mt^N=1.\eeq
}

{
In this context, the scalar product is transformed into a finite-dimensional discrete orthogonality relation with a weight that includes a regularized version of the   ground-state wave function $\sqrt{\Delta_N}$.
The eigenfunctions of the model are still the Macdonald superpolynomials --
thanks to a remarkable symmetry property -- but now corresponding to a finite set of partitions in one-to-one correspondence with the lattice points. The bottom line of this truncation procedure is the finiteness Hilbert space.     
    The  details of this construction will be presented elsewhere.}\\

{\n {\bf{Acknowledgments.}}}  
The authors thank L.~Lapointe for {useful} discussions and for his collaboration on the  article \cite{BDLM2}, on which  part of the present one is based. {We are also grateful to J.F.~van Diejen for a clear and stimulating presentation of his work \cite{vDV}.
Finally, we must thank warmly the anonymous referee B for his/her very useful comments and suggestions which have reshaped our original presentation.}This work was  supported by NSERC, {FRQNT}, FONDECYT  \#1131098.  P.D. is grateful to D.~C\^ot\'e (CRIUSMQ)   for financial support.

\end{document}